\documentclass[preprint,12pt]{elsarticle}

\usepackage{amssymb}
\usepackage{amsmath}
\usepackage{tabularx}
\usepackage{stackengine}
\usepackage{xcolor}
\usepackage{soul}

\journal{Journal of computational physics}

\begin{document}

\begin{frontmatter}



\title{Spectral element lattice Boltzmann method for non-ideal gases with partial wetting boundary condition     }


\author[a]{Chunheng Zhao}
\author[b]{Saumil Patel}
\author[a]{Taehun Lee}

\address[a]{Department of Mechanical Engineering, City College of New York}
\address[b]{Computational Science Division, Argonne National Laboratory}

\begin{abstract}
We present a spectral element lattice Boltzmann method (LBM) for partial wetting on curved geometries. A non-ideal gas phase-field model is incorporated into the LBM framework to enable phase separation with a constant interface thickness and the potential form of surface tension force is used. We adopt the force-splitting approach, yielding significantly improved stability and accuracy. Complex boundaries are naturally handled using a flux bounce-back scheme, which resolves inconsistencies in normal vectors across adjacent elements. Additionally, a general wetting boundary condition is implemented to capture static contact line in a thermodynamically consistent manner. The method is validated through simulations of droplets on flat surfaces, 2/3-dimensional curved surfaces, and equilibrium droplets without boundaries. Results demonstrate that parasitic currents are significantly reduced on unstructured meshes with complex geometries, reaching residual kinetic energy levels on the order of $10^{-24}$ for wetting configurations and $10^{-30}$ for isolated droplets.

\end{abstract}


\begin{highlights}
\item Two-phase spectral Element LBM
\item Wetting in unstructured meshes
\item Force splitting scheme to eliminate parasitic currents
\end{highlights}

\begin{keyword}
Two-phase spectral element LBM \sep General wetting boundary condition \sep Force splitting 
\PACS 0000 \sep 111 1
\MSC 0000 \sep 1111
\end{keyword}

\end{frontmatter}


\section{Introduction}

The lattice Boltzmann method (LBM) is a powerful and efficient approach for simulating two-phase flows, particularly those involving wetting phenomena~\cite{yan2007lattice,lee2008wall,baroudi2020effect,briant2004lattice,zhao2023engulfment}. Among the various formulations, the non-ideal gas model combined with the Cahn–Hilliard free energy has proven effective for modeling two-phase systems where surface tension arises from the Korteweg stress~\cite{zhao2023general,van1979thermodynamic,jacqmin1996energy,lee2006eliminating,nadiga1995investigations}. When expressed in potential form and combined with isotropic finite difference discretizations, this model can eliminate parasitic currents on structured meshes~\cite{lee2008wall,zhao2023general}. Coupling with wetting boundary conditions~\cite{lee2008wall,jacqmin2000contact} further enables accurate capture of contact line dynamics.

A major limitation of the standard LBM is its reliance on integer lattice grids, making it challenging to simulate flows in complex or unstructured geometries~\cite{zhao2025imexlbm,zhao2025difference}. While extensions such as the finite volume LBM~\cite{xi1999finite,mishra2007solving}, finite difference LBM ~\cite{mei1998finite,guo2003explicit,junk2001finite}, and finite element LBM~\cite{lee2001characteristic,min2011spectral,patel2016new,matin2018finite} provide greater mesh flexibility, applying these to two-phase wetting problems introduces two fundamental challenges: (i) the isotropic character of the surface tension discretization cannot be maintained on unstructured meshes, leading to persistent parasitic currents, and (ii) boundaries on unstructured meshes are generally curved or irregular, requiring consistent treatment of normal vectors and thermodynamically consistent enforcement of contact angles across element interfaces.

Building upon our previous work on the spectral element lattice Boltzmann method (LBM) for single-phase flow~\cite{zhao2026flux}, we address these challenges by extending the framework to two-phase flow with partial wetting. The key contributions are: (1) a force-splitting strategy that decomposes the surface tension force into leading- and higher-order components for balanced discretization on unstructured meshes~\cite{patel2016new}, (2) a flux bounce-back scheme that resolves normal vector inconsistencies across contiguous element interfaces while conserving mass and enforcing no-slip conditions~\cite{min2011spectral}, and (3) a thermodynamically consistent wetting boundary condition that couples the wall free energy with the spectral element weak formulation through the density gradient at the contact line~\cite{zhao2023general}.

In the following sections, we present the methodology with emphasis on the coupling between the wetting boundary condition and the spectral element framework (Section 2), followed by validation tests ordered by the primary contribution: wetting on flat surfaces (Section 3.1), wetting on curved surfaces (Section 3.2), and a baseline force-splitting verification using an isolated droplet (Section 3.3).

\section{Methodology}
\subsection{Single distribution function two-phase lattice Boltzmann method}
The single distribution function $f_\alpha$, associated with the fluid density $\rho$, is constructed to model two-phase flow dynamics. The density field is initialized using a hyperbolic tangent profile (Eq. A.1 in Appendix A). The evolution of $f_\alpha$ is governed by the discrete velocity Boltzmann equation:

\begin{equation}\label{dbe_m}
    \left(\frac{\partial}{\partial t}+\boldsymbol{e}_\alpha\cdot\nabla\right)f_\alpha=-\frac{1}{\lambda}(f_\alpha-f_\alpha^{eq})+F_\alpha,
\end{equation}
where $\boldsymbol{e}_\alpha$  denotes the discrete velocity, $\lambda$ is the relaxation time,  and $f_\alpha^{eq}$ represents the equilibrium distribution function (see Appendix A for the full expression). For the current validation cases, we adopt the standard D2Q9 model~\cite{zhao2023interaction}.

Two-phase flow is modeled using a single-distribution function by replacing the ideal gas pressure with a pressure derived from the Cahn–Hilliard free energy. The standard formulation for the bulk free energy, surface energy, and their associated parameters $\beta$, $\kappa$, $\gamma$, follows~\cite{zhao2023general,jacqmin1996energy,nadiga1995investigations} and is summarized in Appendix A (Eqs. A.2–A.7). The key quantity for the force formulation is the chemical potential:
\begin{equation}\label{chemi}
    \mu=\beta(\rho-\rho_l)(\rho-\rho_v)(2\rho-\rho_l-\rho_v)-\kappa\nabla^2\rho.
\end{equation}
The surface tension force is obtained from the potential form $\rho\nabla\mu$~\cite{lee2006eliminating,zhao2023interaction}, yielding the forcing term:
\begin{equation}
   F_\alpha=\frac{t_\alpha}{c_s^2} \left[(\boldsymbol{e}_\alpha-\boldsymbol{u})+\frac{(\boldsymbol{e}_\alpha\cdot\boldsymbol{u})\boldsymbol{e}_\alpha}{c_s^2}\right]\cdot\boldsymbol{F},
\end{equation}
where $\boldsymbol{F}=\nabla\rho c_s^2-\rho\nabla\mu$ and $c_s=1/\sqrt{3}$ represents the speed of sound in two dimensional simulation. The weight function for $\alpha$ direction, $w_\alpha$, follows the regular Lattice Boltzmann method can be found in \cite{zhao2023interaction}. This forcing term is decomposed into a leading-order term $F_\alpha^*=\frac{t_\alpha}{c_s^2}\boldsymbol{e}_\alpha\cdot\boldsymbol{F}$ and a higher-order term $F_\alpha^{**}=\frac{t_\alpha}{c_s^2}
\left[\frac{(\boldsymbol{e}_\alpha\cdot\boldsymbol{u})\boldsymbol{e}_\alpha}{c_s^2}-\boldsymbol{u}\right]\cdot\boldsymbol{F}$ involving velocity-dependent corrections~\cite{patel2016new}. This decomposition is central to achieving balanced discretization, as detailed below.

The collision step incorporates the higher-order forcing term:
\begin{equation}
    f_\alpha^*=\bar{f}_\alpha-\frac{1}{\tau+0.5}(\bar{f}_\alpha-\bar{f}_\alpha^{eq})+\Delta t F_\alpha^{**},
\end{equation}
where the modified distribution functions absorb the higher-order correction (see Appendix A). The streaming step is formulated as a weak problem over each spectral element $\Omega_e$:
\begin{equation}\label{advection}
    \left(\frac{\partial\bar{f}_\alpha}{\partial t }+\boldsymbol{e}_\alpha\cdot\nabla \bar{f}_\alpha -F_\alpha^*,\phi\right)_{\Omega_e}=\left(J_\alpha,\phi\right)_{\partial\Omega},
\end{equation}
where the right-hand side contains the boundary flux $J_\alpha$ from the flux bounce-back scheme and $\phi$ is the test function~\cite{zhao2026flux}. The $\partial\Omega$ denotes the boundary. In matrix form:
\begin{equation}\label{spetrum}
    \boldsymbol{M}\frac{\partial\bar{f}_\alpha}{\partial t }+\boldsymbol{C}_\alpha \bar{f}_\alpha =\boldsymbol{M}F_\alpha^*+\boldsymbol{R}\boldsymbol{J}_\alpha,
\end{equation}
where $\boldsymbol{M}$ and $\boldsymbol{C_\alpha}$ are the mass and convection matrices respectively, and $\boldsymbol{R}$ is the surface integration. The derivations of all the matrices mentioned above can be found in our previous paper~\cite{zhao2026flux}. Time integration is performed using the third-order strong stability preserving Runge–Kutta scheme~\cite{zhao2026flux}.

\subsection{Thermodynamically Consistent Wetting Boundary Conditions}

To incorporate the wetting effect and enforce the correct contact-angle boundary condition, a wall free-energy density $e_w$ is introduced following the thermodynamically consistent formulations in \cite{zhao2023general,liu2009wall}:
\begin{equation}
e_w=\cos{\theta^{eq}}\int_{\rho_{vs}}^{\rho}\sqrt{2\kappa\rho e_0} d\rho,
\end{equation}
where $\rho_{vs}$ represents the density of the solid gas surface, aligning with the equilibrium vapor density, $\rho_{vs}=\rho_v$. 

The enforcement of the contact angle enters the spectral element formulation through two coupled mechanisms:
(1) Density gradient at the wall. In equilibrium, the wall free energy yields the relation $\partial_n\rho=\cos\theta^{eq}\sqrt{2 e_0/\kappa}$, where $\partial_n\rho$ is the density gradient normal to the wall. This condition specifies the interface orientation at the contact line, ensuring that the prescribed contact angle is maintained with a consistent interface thickness. (2) Coupling through the Laplacian in weak form. The chemical potential (Eq.~\ref{chemi}) contains the term $
\kappa\nabla^2\rho$, which in the spectral element framework is evaluated via the weak form:
\begin{equation}\label{weak}
\nabla^2\rho = \boldsymbol{M}^{-1} \left( \int_{\partial\Omega} (\phi \boldsymbol{n} \cdot \nabla\rho) d\bar{\Omega} - \int_\Omega (\nabla\rho \cdot \nabla\phi) d\Omega \right).
\end{equation}

The boundary integral in Eq.~\ref{weak} directly incorporates the wetting condition: $\boldsymbol{n} \cdot \nabla\rho = \cos\theta^{eq}\sqrt{2e_0/\kappa}$. Through this mechanism, the contact angle information propagates into the chemical potential $\mu$, which in turn modifies the leading-order forcing term $F^*_\alpha$ in the streaming equation (Eq.~\ref{advection}). Thus, the wall free energy modifies the force balance in streaming—not the distribution function fluxes $J_\alpha$ directly. The flux bounce-back (Eq. 9 below) enforces only the no-slip condition and mass conservation, while the contact angle is enforced through the modified chemical potential gradient.
\begin{figure}
	\centering
  \includegraphics[width=0.9\linewidth]{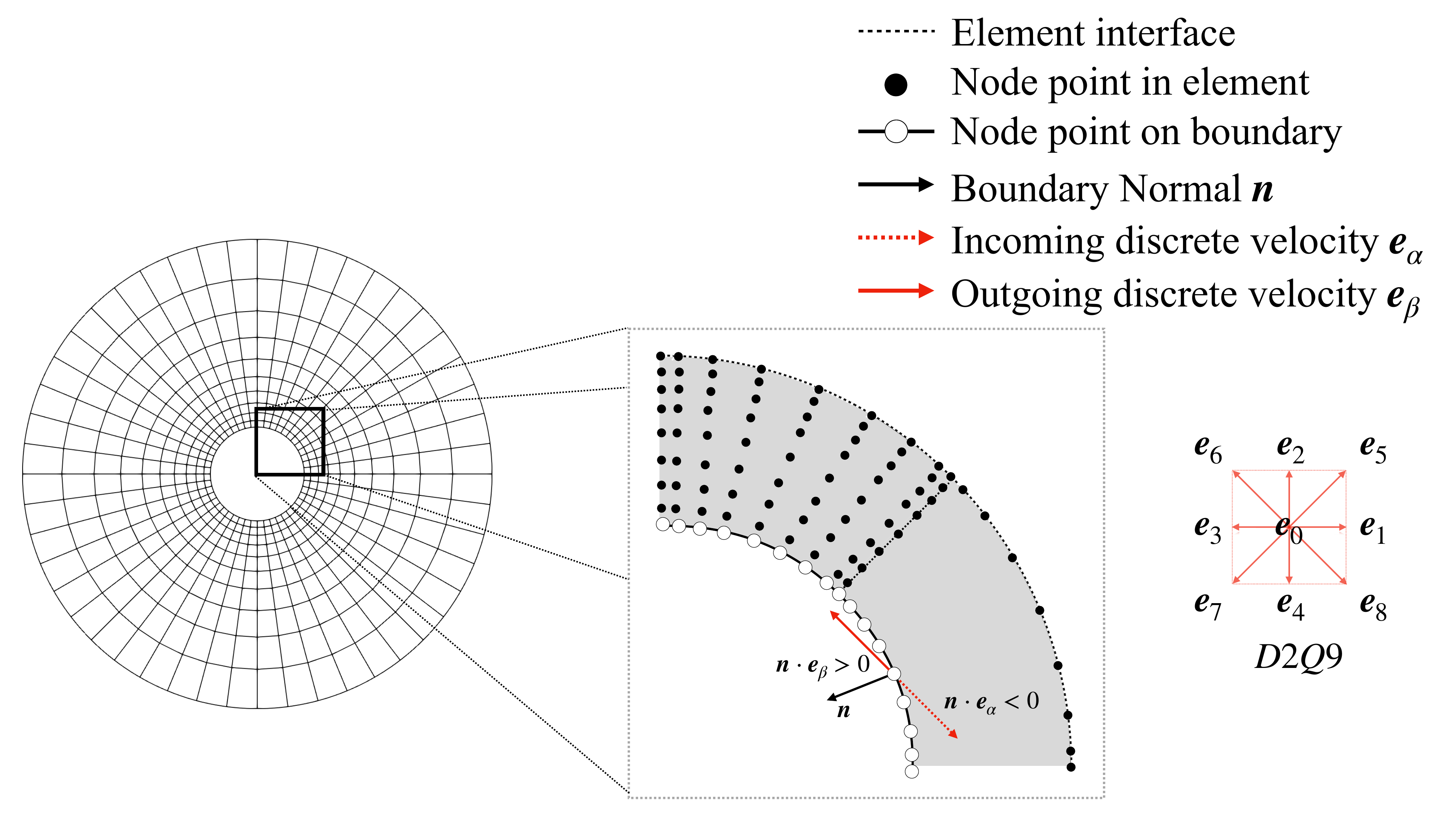}
    \caption{\label{wall} Schematic of the flux bounce-back scheme for a boundary node point on the curved boundary with polynomial order $N=8$. The incoming direction $\alpha$ and the corresponding bounce-back (outgoing) direction $\beta$ are defined with respect to the boundary normal $\boldsymbol{n}$. The associated velocity vectors are represented by the red dashed (incoming) and red solid (outgoing) arrows. We indicate the $D2Q9$ model used in our current scheme.
    }
\end{figure} 
The boundary flux is defined as $\boldsymbol{J}_\alpha=\boldsymbol{n}\cdot \boldsymbol{j}_\alpha$. with:
$$ \boldsymbol{j}_\alpha=\left\{
\begin{aligned}
&\boldsymbol{e}_\alpha[\boldsymbol{\bar f}_\alpha]_{bc},&     &\boldsymbol{n}\cdot\boldsymbol{e}_\alpha <0, \\
&\boldsymbol{0} & & \boldsymbol{n}\cdot\boldsymbol{e}_\alpha \ge0.
\end{aligned}
\right.
$$
\begin{equation}
    [\boldsymbol{\bar f}_\alpha]_{bc}=\boldsymbol{\bar f}_\alpha-\boldsymbol{\bar f}_\beta-\frac{2t_\alpha}{c_s^2}\rho(\boldsymbol{e}_\alpha\cdot \boldsymbol{u}_b),
\end{equation}
where $\alpha$ and $\beta$ denote the bounce-back pair directions of the distribution functions (shown in Figure~\ref{wall}), and $\boldsymbol{u}_b$ is the boundary velocity vector. Further implementation details can be found in~\cite{min2011spectral}.

Following the streaming step, the macroscopic variables are updated:

\begin{equation}\label{density}
    \rho=\sum_\alpha \bar{f}_\alpha,
\end{equation}
\begin{equation}\label{momentum}
    \rho\boldsymbol{u}=\sum_\alpha \bar{f}_\alpha \boldsymbol{e}_\alpha.
\end{equation}
Additional details on the spectral element method and the derivation of the mass, convection, and surface operators can be found in ~\cite{deville2002high}.

\section{Numerical Validation}

\subsection{Droplet Wetting on a Flat Surface}

\begin{figure}
	\centering
  \includegraphics[width=\linewidth]{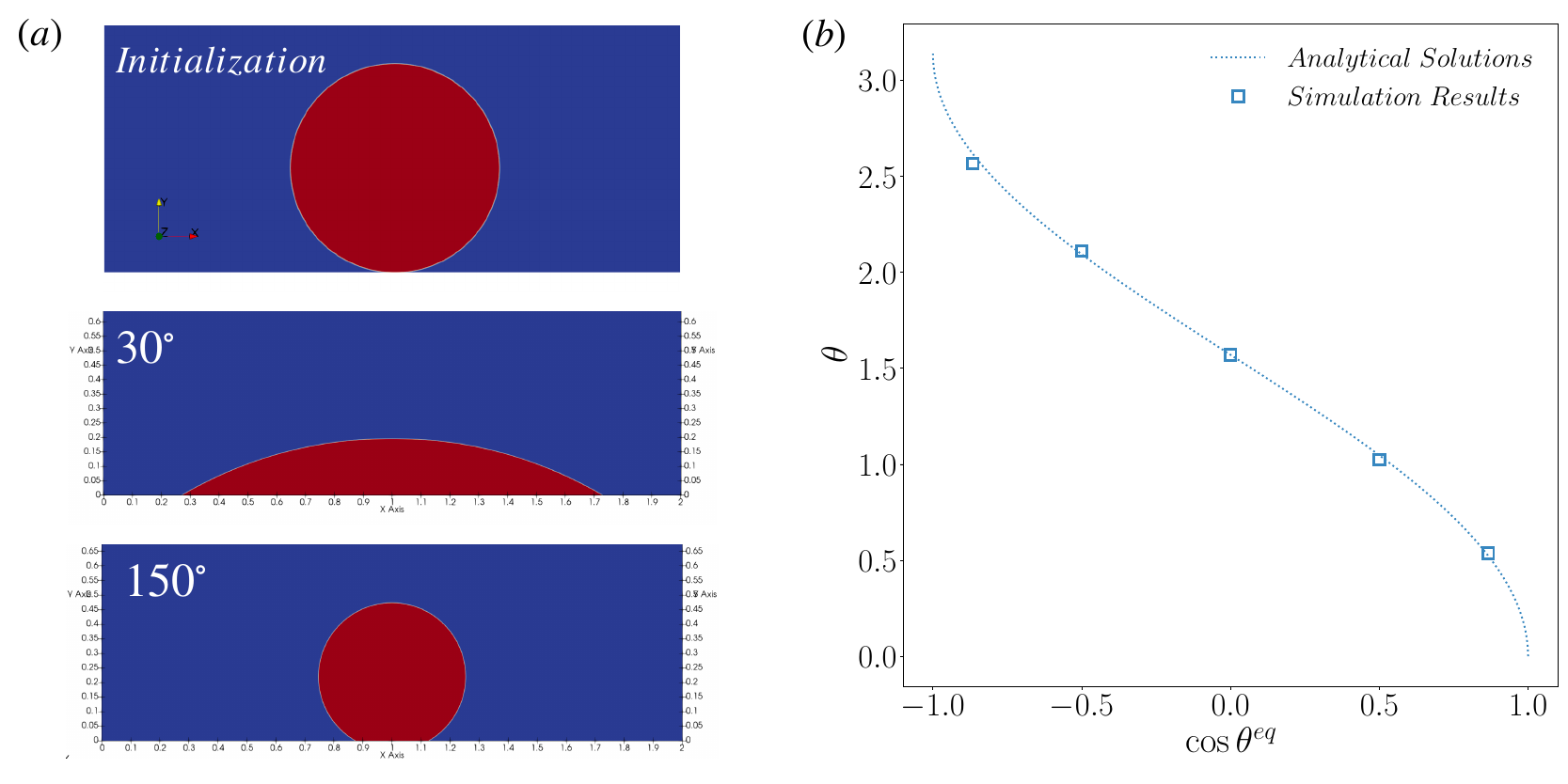}
    \caption{\label{ca} (a) Initialization of the simulation and simulation results for droplet wetting on a flat surface with equilibrium contact angle $\theta^{eq}=\frac{\pi}{6}$ and $\theta^{eq}=\frac{5\pi}{6}$. (b) Comparison between the simulation results with the analytical solutions.
    }
\end{figure}

We first evaluate the proposed scheme by simulating droplets on a flat surface at equilibrium contact angles. A droplet of diameter $D=0.5$, interface thickness $\delta=0.03$, is placed in a rectangular domain $L_1\times L_2=1\times 2$,  positioned such that it just touches a flat substrate (Figure~\ref{ca} (a)). In addition, the mesh is configured with element number for each direction, $N_e=16$, along the $x-axis$ and $N_e=8$ along the $y-axis$. The polynomial order is set to $N=16$. By varying the wall free energy, the system reaches equilibrium contact angles in the range $\theta^{eq}=[\frac{\pi}{6},\frac{5\pi}{6}]$, with the Laplace number fixed at $La=122$. To eliminate the influence of spurious wave reflections, no-slip boundary conditions are applied on the remaining three domain boundaries. The initial density and velocity fields are prescribed as in Eqs.~A.8–A.9 (Appendix A).

\begin{figure}
	\centering
  \includegraphics[width=\linewidth]{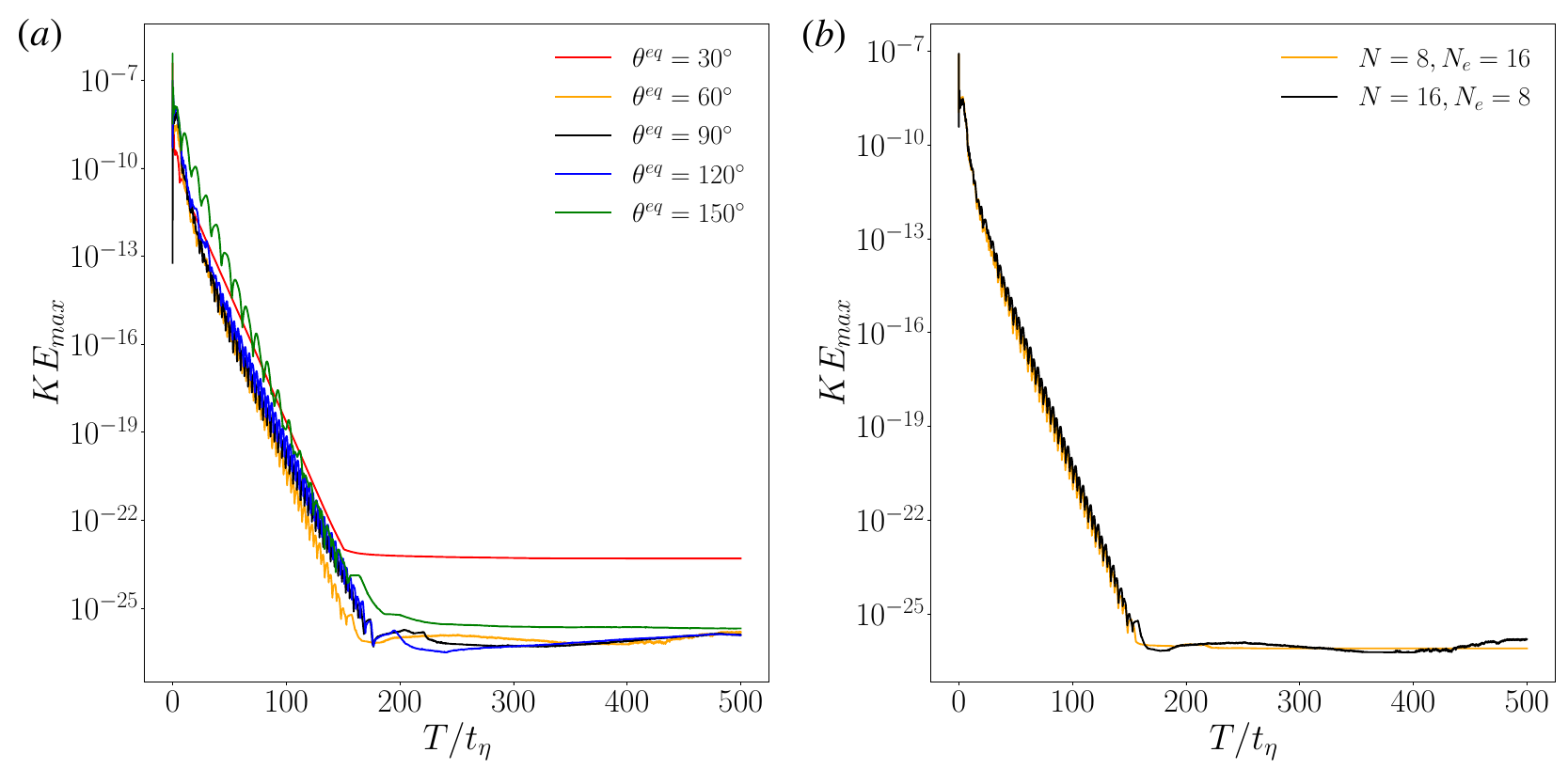}
    \caption{\label{keangle} (a) Maximum kinetic energy $KE_{max}$ evolution for a droplet contact with boundary with equilibrium contact angle $\theta^{eq}=[\frac{\pi}{6},\frac{5\pi}{6}]$, $La=122$, during $T/t_\eta=[0,500]$. (b) Comparison of maximum kinetic energy, $KE_{max}$, for $\theta^{eq}=\frac{\pi}{3}$ with $N=8$, $N_e=16$ and $N=16,N_e=8$. 
    }
\end{figure}
After initialization, the droplet contacts the substrate, and the contact angle evolves toward the equilibrium value implicitly determined by the wall free energy. Once the simulation has run sufficiently long to reach equilibrium, the contact angle is calculated geometrically following the method described in~\cite{zhao2023general}. Figure~\ref{ca} (b) presents a comparison between the simulation results and theoretical predictions, showing excellent agreement with the predefined contact angles.

We further evaluate the parasitic currents, which constitute the central test of the force-splitting scheme in the presence of boundaries. To ensure a consistent analysis across different contact angles, each droplet is initialized with its corresponding equilibrium shape, eliminating additional contact line motion and isolating the parasitic currents.

In this test, we maintain Laplace number, $La=122$. The maximum kinetic energy eventually decreases to $KE_{max}\sim 10^{-27}$ for equilibrium contact angles $\theta^{eq}=[\frac{\pi}{3},\frac{5\pi}{6}]$, and to $KE_{max}\sim10^{-24}$ for $\theta^{eq}=\frac{\pi}{6}$ (shown in Figure~\ref{keangle} (a)). It is noted that a small equilibrium contact angle introduces a relatively large wall free energy source from the boundary to be dissipated by the viscous force. A comparison of $KE_{max}$ for $N=8, N_e=16$ and $N=16, N_e=8$ at the same $La$ and equilibrium contact angle $\theta^{eq}=\pi/3$ is shown in Figure~\ref{keangle} (b). For both polynomial orders, similar parasitic current intensities are observed, with $KE_{max}\sim10^{-27}$.


Upon examination, the residual parasitic currents are found to be asymmetric and of very small magnitude. Unlike regular LBM on structured grids, the spectral element method cannot preserve full isotropy, particularly near wetting boundaries. As a result, the residual kinetic energy in wetting cases ($\sim10^{-24}$ to $10^{-27}$) is higher than in the boundary-free single droplet test ($\sim10^{-30}$, see Section 3.3). This difference of three to six orders of magnitude reflects the inherent anisotropy introduced by boundary discretization in the spectral element framework.  For small contact angles, the liquid vapor interface is positioned very close to the solid surface, hindering the attainment of the elevated equilibrium density near the wall and thus leading to a relatively larger $KE_{max}$~\cite{liu2009wall}. Additionally, at small contact angles, the effective Laplace number $La$, computed from the curvature of the droplet, is approximately 4 to 5 times larger than that of a circular droplet. As a result, achieving parasitic currents of the same order as those observed in other contact angle tests becomes challenging~\cite{zhao2023general}.

\begin{figure}
	\centering  \includegraphics[width=\linewidth]{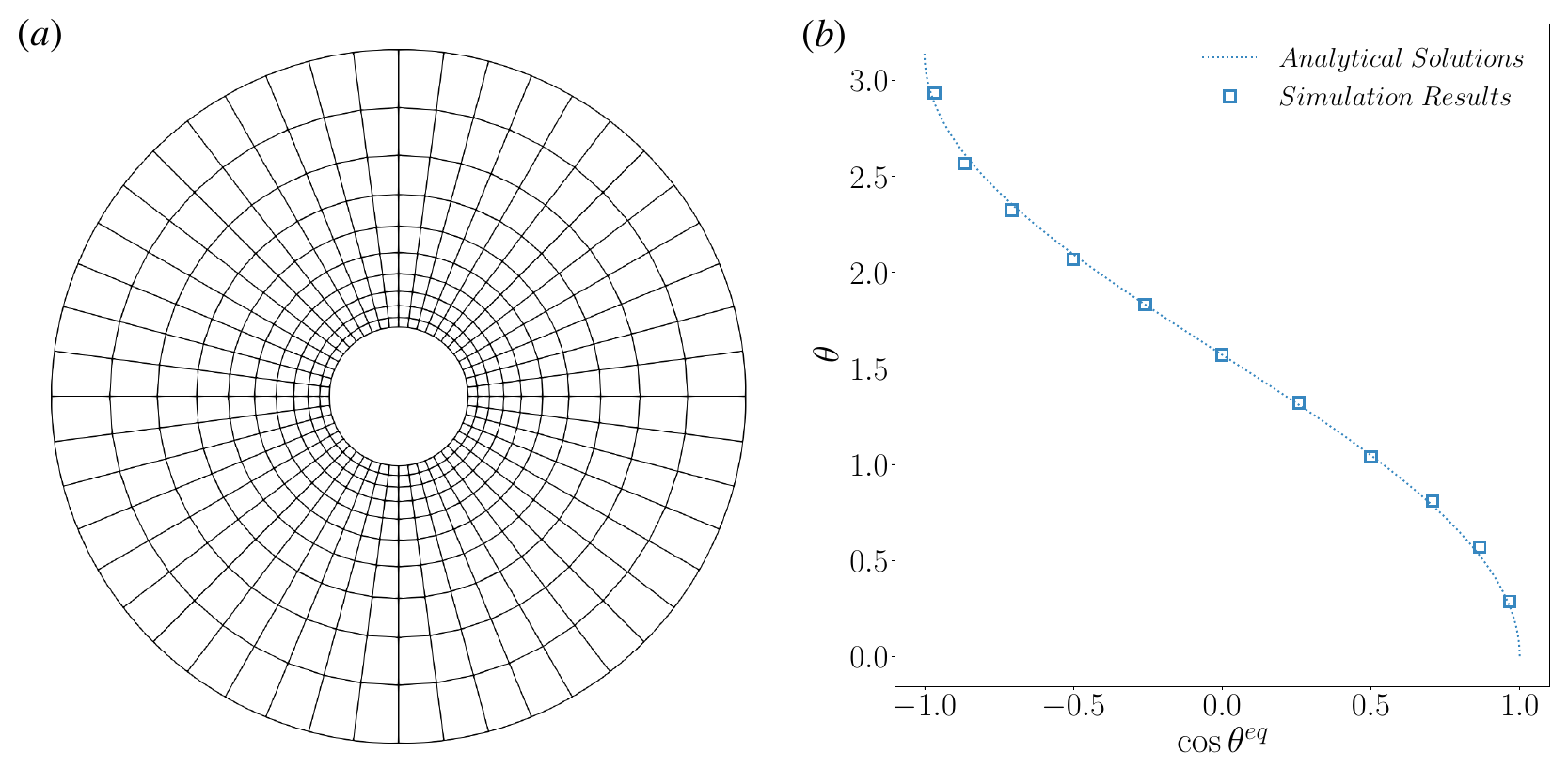}
    \caption{\label{mesh} (a) Mesh structure with a circle boundary in the center. (b) Comparison between the simulation results with the analytical solutions.
    }
\end{figure}
\begin{figure}
	\centering
  \includegraphics[width=\linewidth]{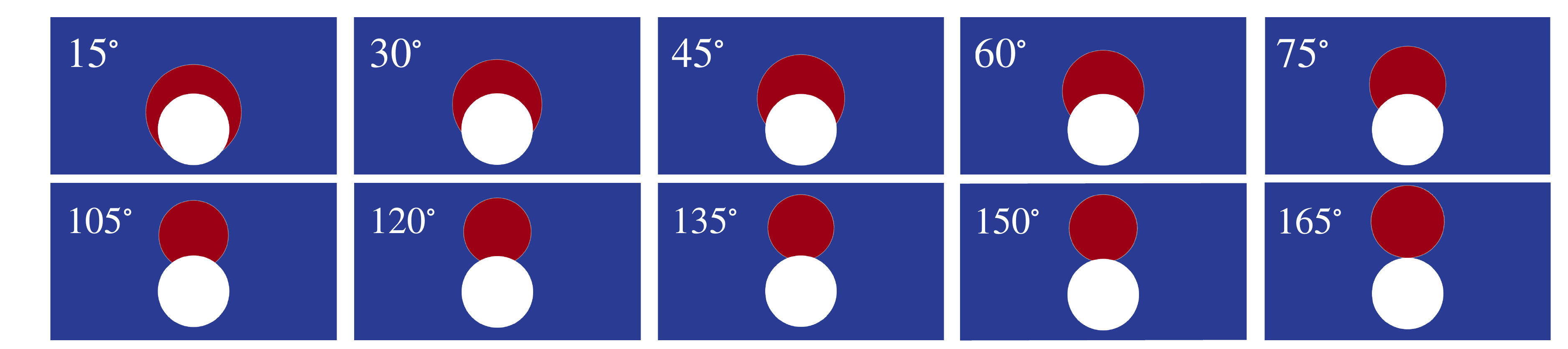}
    \caption{\label{cac} Density profile for droplet wetting on a curved boundary with equilibrium contact angle $\theta^{eq}=[\frac{\pi}{12},\frac{11\pi}{12}]$.
    }
\end{figure}

\begin{figure}
	\centering
  \includegraphics[width=0.5\linewidth]{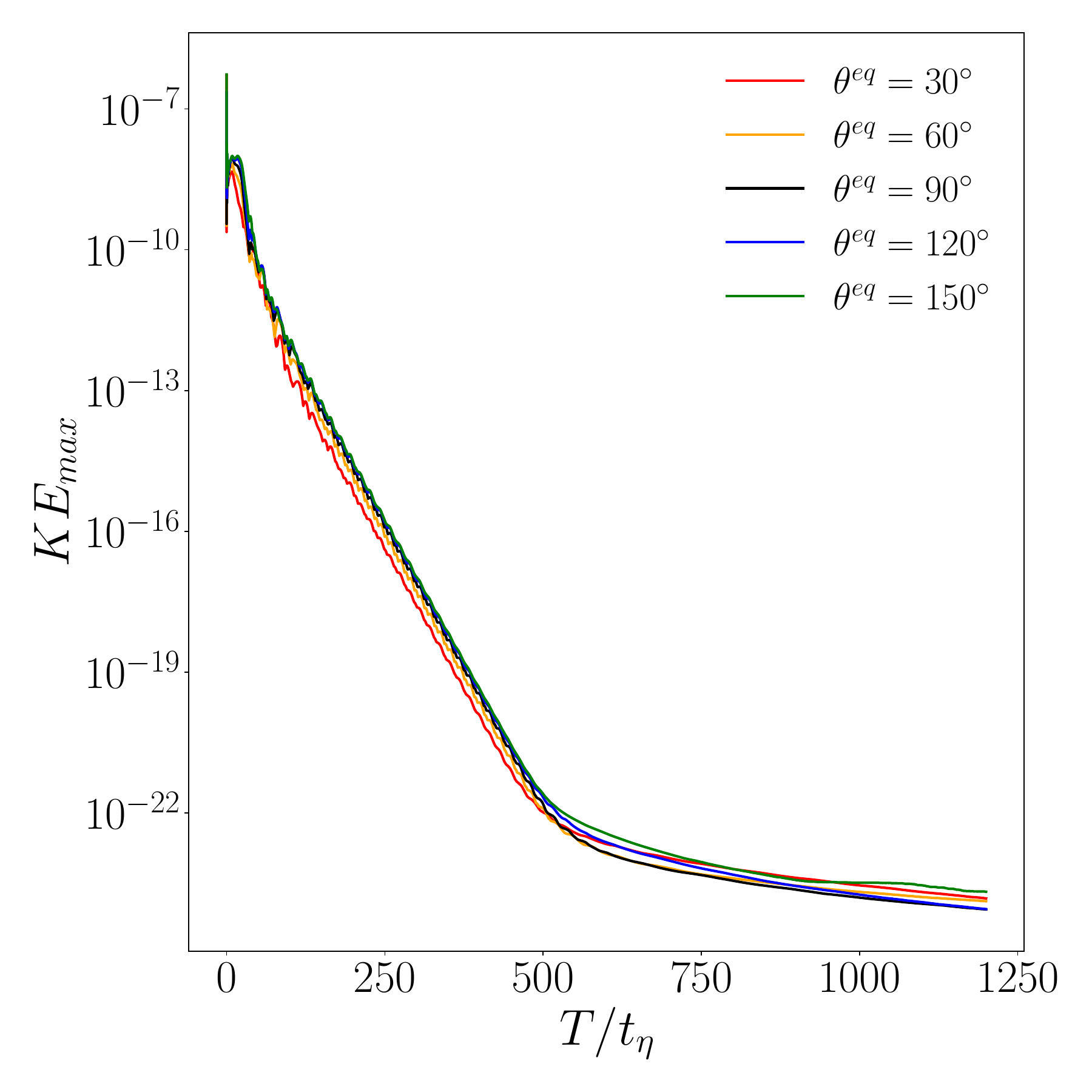}
    \caption{\label{curve} Maximum kinetic energy $KE_{max}$ evolution for the single droplet wetting on the circle boundary with equilibrium contact angle $\theta^{eq}=[\frac{\pi}{6},\frac{5\pi}{6}]$, $La=442$ during $T/t_\eta=[0,1200]$.
    }
\end{figure}

\begin{figure}
	\centering
  \includegraphics[width=\linewidth]{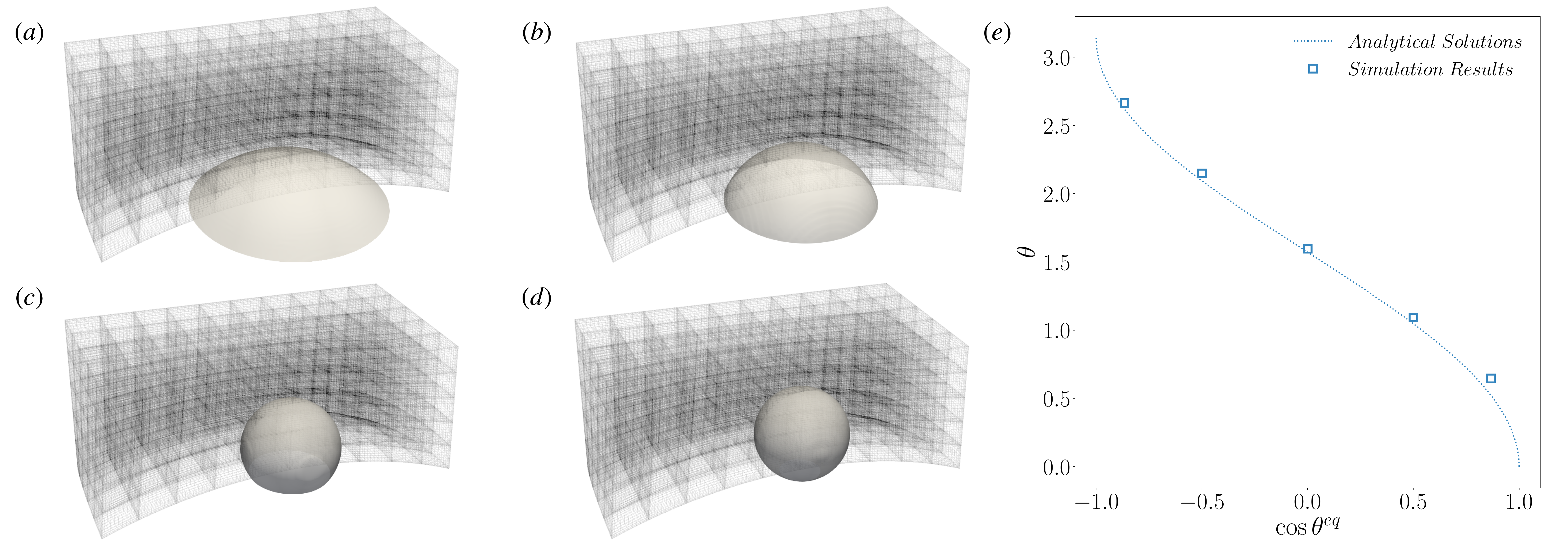}
    \caption{\label{3D} Density contour for $\rho=0.55$ for droplet wetting on a 3D curved surface with equilibrium contact angle (a) $\theta^{eq}=\frac{\pi}{6}$, (b) $\theta^{eq}=\frac{\pi}{3}$, (c) $\theta^{eq}=\frac{2\pi}{3}$, (d) $\theta^{eq}=\frac{5\pi}{6}$. (e) Comparison between the simulation results with the analytical solutions for 3D droplet wetting on a curved surface.
    }
\end{figure}

\subsection{Droplet wetting on a curved surface}
We further simulate droplets on a curved surface at equilibrium contact angles. The mesh is generated using GMSH, as shown in Figure~\ref{mesh} (a), featuring a curved circular boundary with diameter $D=0.2$ located at the center of the 2D square domain. A droplet of the same diameter is initialized just touching this curved boundary, and the wall free energy is varied to simulate contact angles ranging from $\theta^{eq}=[\frac{\pi}{12},\frac{11\pi}{12}]$. In this test, we set $La=341$, and the polynomial order is set to $N=11$.

The simulation results are obtained after long-time evolution to ensure equilibrium is reached. In Figure~\ref{mesh} (b), we compare the simulation results (indicated by square markers) with the analytical solution, showing excellent agreement. Additional comparisons for each equilibrium contact angle are presented in Figure~\ref{cac}.

To evaluate parasitic currents for simulations with different contact angles  $\theta^{eq}=[\frac{\pi}{6},\frac{5\pi}{6}]$, we initialize the droplet on the surface directly with the equilibrium contact angle. All simulations exhibit a consistent decreasing trend in kinetic energy, with $KE_{max}\sim10^{-24}$. Our previous work~\cite{patel2016new} indicates that the parasitic currents are related to the mesh structure. A perturbed mesh will induce a relative strong parasitic currents. Although the kinetic energy continues to decline over time, the process is extremely slow, consistent with the results in~\cite{patel2016new}. 

A 3D study was performed to investigate droplet wetting on a curved surface, with equilibrium contact angles ranging from  $\pi/6$ to $5\pi/6$. The computational domain is a rectangular cuboid ($L=0.5, W=0.5, H=0.25$) featuring a curved bottom boundary with a diameter of $D=1.4$. A droplet with a diameter of $d=0.25$ was initialized at $(0.25, 0.25, 0.04)$ in contact with the spherical surface. The D3Q13 lattice model  \cite{shan2006kinetic} was employed to ensure spatial isotropy while maintaining computational efficiency. 

The simulation parameters ($Cn = \delta/d = 0.08$, $La = 60$, $CFL=0.75$) differ from the 2D cases due to the computational cost of 3D simulations: the lower Laplace number and larger Cahn number reduce the required spatial and temporal resolution while still providing a meaningful validation of contact angle accuracy. After $T/t_\eta \approx 60$, parasitic currents decreased to $\sim10^{-11}$. Although these currents were still gradually decaying, the results demonstrate excellent agreement with the prescribed equilibrium contact angles, as illustrated in Figure~\ref{3D}.

\subsection{Baseline Force-Splitting Verification: Static Droplet}
\begin{figure}
	\centering
  \includegraphics[width=\linewidth]{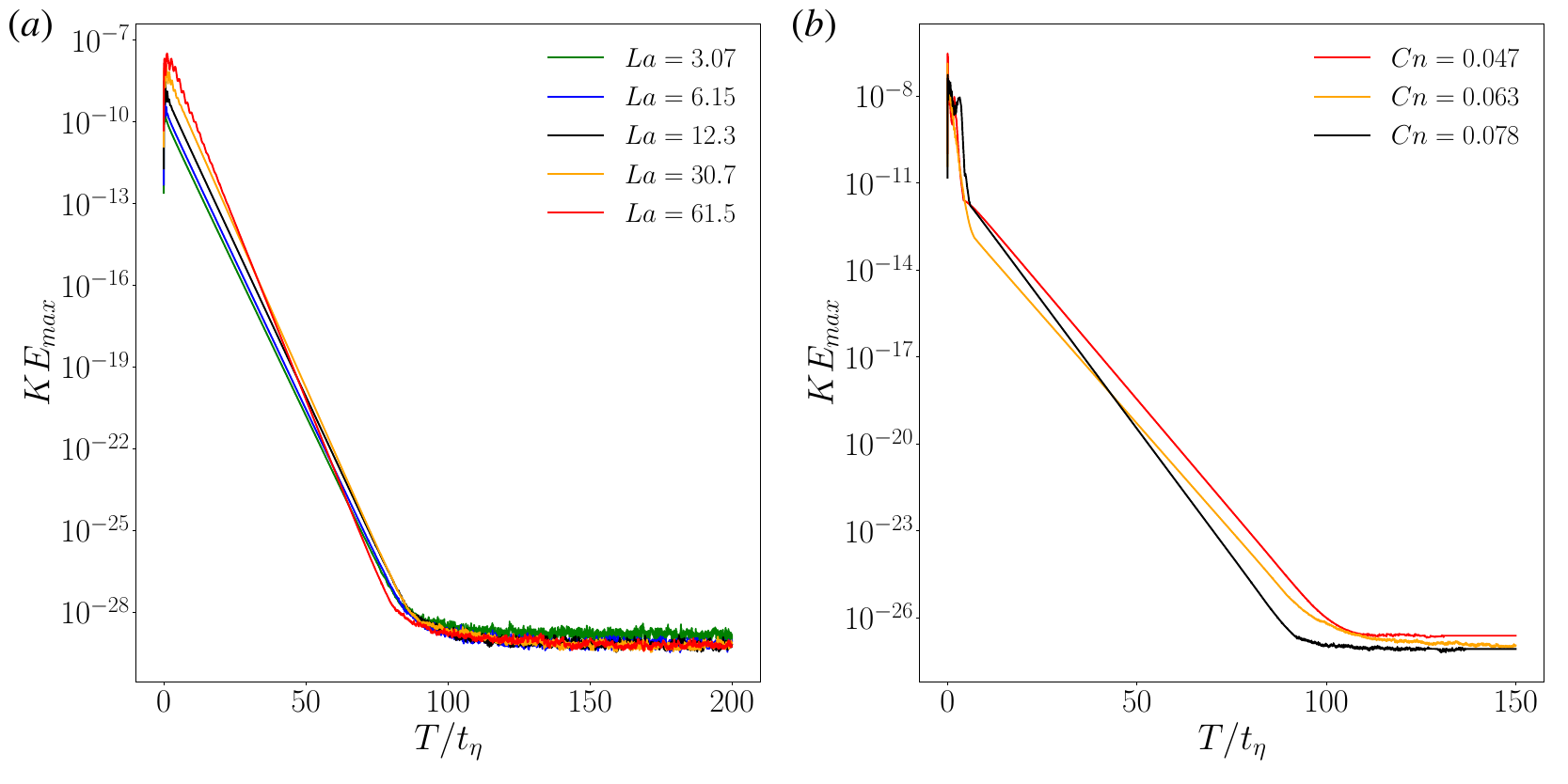}
    \caption{\label{para} Maximum kinetic energy $KE_{max}$ evolution for the single droplet with (a) $La=[3.07,61.5]$ during $T/t_\eta=[0,200]$; (b) $Cn=[0.047,0.078]$ during $T/t_\eta=[0,150]$.
    }
\end{figure}
In regular LBM, using the potential form of the surface tension force and isotropic finite difference, we can eliminate parasitic currents, as demonstrated in~\cite{lee2006eliminating}. For the spectral element LBM, the streaming is solved by the Runge–Kutta method, and we cannot maintain the isotropic character when incorporating the surface tension force in collision step. Nevertheless, parasitic currents can be further reduced by incorporating the leading-order forcing term into the streaming step. Following the force-splitting approach~\cite{patel2016new}, the leading-order contribution of the forcing term is consistently applied during streaming:
\begin{equation}
 \frac{\partial\bar{f}_\alpha}{\partial t }+\boldsymbol{e}_\alpha\cdot\nabla \left(\bar{f}_\alpha-t_\alpha\rho\right) -\frac{t_\alpha}{c_s^2}\rho\boldsymbol{e}_\alpha\cdot\nabla\mu=0.
\end{equation}
As the system approaches equilibrium, the second term vanishes, and the chemical potential generates an isotropic force along the interface, leading to a balanced system.

A droplet of diameter $D=0.5$ is initially placed at the center of a square computational domain of size $L=1$. Simulations are performed for $La=[3.07,61.5]$, where $La=\gamma D/\eta^2$ with the relaxation time fixed at  $\tau=0.5$ for both phases, over  $T/t_\eta=[0,200]$, where $t_\eta=\eta D/\gamma$ represents the viscous time scale.

As shown in Figure~\ref{para} (a), $KE_{max}=0.5\rho\boldsymbol{|u|}^2_{max}$ is monitored over the dimensionless time interval $T/t_\eta=[0,200]$ till the system reaches equilibrium. To ensure consistency, the Cahn number is fixed at $Cn=\delta/D=0.063$ for all simulations. In this test, the element count in each spatial direction is set to $N_e=4$, and the polynomial order is $N=16$. The results show that $KE_{max}$ decays exponentially during the initial stage, up to approximately $T/t_\eta\sim100$. Droplets with larger $La$ exhibit a steeper decline, indicating faster relaxation toward equilibrium compared to those with smaller $La$. At longer times, all cases converge to a residual kinetic energy of approximately $KE_{max}\sim10^{-30}$, demonstrating that the system reaches a well-established equilibrium state. For comparison, Figure \ref{para} (b) presents the evolution of $KE_{max}$ for $N_e=16$, and $N=4$ (corresponding to the same total number of grid points) with $Cn=[0.047,0.078]$ over $T/t_\eta\sim150$. In all cases, the bulk energy coefficient is kept constant at $\beta=0.001$. As the polynomial order decreases, the lower polynomial order leads to a higher residual kinetic energy, converging to $KE_{max}\sim10^{-28}$.

This baseline establishes the best achievable residual kinetic energy for the spectral element framework in the absence of boundaries. The subsequent introduction of wetting boundaries (Sections 3.1–3.2) raises the residual to $\sim10^{-24}$ due to boundary-induced anisotropy, providing a clear quantification of the boundary effect on parasitic currents.

\section{Concluding remarks}
A well-balanced spectral element lattice Boltzmann method (LBM) is developed for simulating static partial wetting on unstructured meshes. The method integrates the non-ideal gas LBM model with a spectral element framework~\cite{min2011spectral,patel2016new,zhao2023general} and is employed to examine parasitic currents in three representative configurations: a single droplet, a droplet wetting a flat surface, and a droplet wetting a curved surface. A force-splitting strategy is adopted to add the forcing term consistently at each Runge–Kutta stage~\cite{zhao2026flux}. When coupled with the potential form of the surface-tension force, this method significantly reduces parasitic currents. 

For the boundary-free single droplet test, the force-splitting scheme achieves residual kinetic energy at $KE_{max}\sim 10^{-30}$, demonstrating effective elimination of parasitic currents. In wetting configurations, boundary-induced anisotropy inherent to the spectral element discretization raises residuals to $\sim10^{-24}$ to $\sim10^{-27}$, depending on the contact angle and geometry. While these residuals are significantly lower than those typically observed in standard unstructured LBM approaches, they reflect a fundamental trade-off: the spectral element method cannot fully preserve isotropy near boundaries, particularly in wetting problems where the liquid–vapor interface lies close to the solid surface. This trade-off is most pronounced at small contact angles, where the effective Laplace number increases substantially.

In this study, no filtering is applied, unlike the standard practice in spectral element methods for turbulent flows~\cite{deville2002high}. The application of filters can interfere with the force balance, either by altering the macroscopic quantities (density/velocity) or by affecting the distribution functions. However, our current formulation is restricted to Van der Waals fluids. While the model allows for large density ratios, the evolution remains slow due to low Mach number constraints. Under these conditions, filtering is not necessary for maintaining stability. Future work will focus on extending the method to incompressible flows with larger characteristic velocities, where filter effects must be carefully evaluated across different scenarios.
\section{acknowledgment}
This material is based upon work supported by the U.S. Department of Energy (DOE), Office of Nuclear Energy, under Award No. DE-NE0009420, and the National Science Foundation under Grant No. 2344147. This research used resources of the Argonne Leadership Computing Facility, which is a U.S. Department of Energy Office of Science User Facility operated under contract DE-AC02-06CH11357.

\appendix
\section{Standard Free Energy Formulations }
\label{sec:Appendix}

The density field is initialized with a two-phase profile:
\begin{equation}\label{dens}
        \rho=\frac{\rho_l+\rho_v}{2}-\frac{\rho_l-\rho_v}{2}\tanh\left({\frac{2\boldsymbol{x}}{\delta}}\right),
\end{equation}
where $\boldsymbol{x}$ represents the signed distance to the  interface, and $\delta$ is the constant interface thickness. In most of our simulations, $\rho_l=1$ and $\rho_v=0.1$ are used. Higher density ratios may also be considered, as demonstrated in previous studies~\cite{lee2006eliminating}.

The equilibrium distribution function is given by:
\begin{equation}\label{geq_s}
    f_\alpha^{eq}=\rho t_\alpha\left(1+\frac{\boldsymbol{e}_\alpha \cdot \boldsymbol{u}}{c_s^2}+
\frac{(\boldsymbol{e}_\alpha\cdot\boldsymbol{u})^2}{2c_s^4}-
\frac{|\boldsymbol{u}|^2}{2c_s^2}\right).
\end{equation}
The total free energy of the system is:
\begin{equation}
    E=\int_\Omega \left(e_0+e_s\right)d \Omega+\int_{\partial\Omega}e_wd\partial\Omega,
\end{equation}
where the bulk free energy density is $e_0=\beta(\rho-\rho_l)^2(\rho-\rho_v)^2$, the surface energy density is $e_s=0.5\kappa|\nabla\rho|^2$, and $\beta$, $\kappa$ denote the  bulk energy coefficient and surface tension coefficient respectively. The parameter relations are:
\begin{equation}
    \kappa=\frac{\beta\delta^2(\rho_l-\rho_v)^2}{8},
\end{equation}
\begin{equation}
    \gamma=\frac{(\rho_l-\rho_v)^3}{6}\sqrt{2\kappa\beta},
\end{equation}
The modified distribution functions used in the collision step are:
\begin{equation}
    \bar{f}^{eq}_\alpha=f^{eq}_\alpha-0.5\Delta t F^{**}_\alpha,
\end{equation}
\begin{equation}
    \bar{f}_\alpha=f_\alpha-0.5\Delta t F^{**}_\alpha.
\end{equation}
The initial conditions for the wetting simulations:
\begin{equation}
    \rho(\boldsymbol{x},0)=\frac{\rho_l+\rho_v}{2}-\frac{\rho_l-\rho_v}{2}\tanh\left({\frac{2\left(|\boldsymbol{x}-\boldsymbol{x_0}|-R\right)}{\delta}}\right).
\end{equation}
\begin{equation}
    \boldsymbol{u}(\boldsymbol{x},0)=\boldsymbol{0}.
\end{equation}

 \bibliographystyle{elsarticle-num} 
 \bibliography{cas-refs}





\end{document}